\documentclass[pra,twocolumn,showpacs,aps,amssymb,superscriptaddress]{revtex4-1}
\newcommand{\Eref}[1]{Eq. (\ref{#1})}

\newcommand{\Fref}[1]{Fig. \ref{#1}}
\newcommand{\Sref}[1]{Sec. \ref{#1}}
\newcommand{\tr}{\mathrm{tr}}

\newcommand{\ket}[1]{{| #1 \rangle }}
\newcommand{\ketbra}[2]{{| #1 \rangle \langle #2 |}}

\usepackage[colorlinks,bookmarks=false,citecolor=blue,linkcolor=red,urlcolor=blue]{hyperref}
\usepackage{graphicx}
\usepackage{amsmath}
\usepackage{amssymb}
\usepackage{wasysym}
\usepackage{verbatim}
\usepackage[usenames,dvipsnames]{xcolor}
\usepackage{times}
\usepackage{color}
\usepackage{grffile}
\usepackage{color}

\begin{document}

\title{Dynamical and Steady State Properties of a Bose-Hubbard Chain with Bond-Dissipation:\\A Study based on Matrix Product Operators}

\author{Lars Bonnes}
\email{lars.bonnes@uibk.ac.at}
\address{Institute for Theoretical Physics, University of Innsbruck, A-6020 Innsbruck, Austria}

\author{Daniel Charrier}
\address{Max-Planck-Institut f\"ur Physik komplexer Systeme, N\"othnitzer Str. 38, 01187 Dresden, Germany}

\author{Andreas M. L\"auchli}
\address{Institute for Theoretical Physics, University of Innsbruck, A-6020 Innsbruck, Austria}

\date{\today}

\begin{abstract}
We study a dissipative Bose-Hubbard chain subject to an engineered bath using a superoperator approach based 
on matrix product operators. The dissipation is engineered to stabilize a  BEC condensate wavefunction in its steady
state. We then characterize the steady state emerging from the interplay between incompatible Hamiltonian
and dissipative dynamics. While it is expected that interactions lead to this competition, even the kinetic energy in an 
open boundary condition setup competes with the dissipation, leading to a non-trivial steady state. We also present results 
for the transient dynamics and probe the relaxation time revealing the closing of the dissipative gap in the thermodynamic limit.
\end{abstract}
\maketitle

\section{Introduction}
The preparation of quantum states in standard cold-atom experiments relies on the ability to cool the system to extremely low temperatures and then transfer the isolated system adiabatically to the target state~\cite{bloch07}.
Dissipative state preparation, on the other hand, pursues a different route towards the realization of complex quantum states.
By coupling the system to a suitably designed bath, a non-unitary time evolution will drive the system into a (unique) pure steady-state that has the desired properties such as long-range phase coherence~\cite{diehl08,kraus08}.
This method is particularly appealing since the steady-state is often an attractor for the time evolution of the open systems for almost arbitrary initial states.
Recent proposals address a wide range of applications to states with long range phase coherence~\cite{diehl08,kraus08}, matrix product states~\cite{kraus08}, Kitaev wires~\cite{diehl11b} or $p$-wave superfluids with Majorana edge modes~\cite{bardyn12} and other topologically non-trivial phases~\cite{bardyn13}.
Recent experiments with trapped ions demonstrate the feasibility of this concept of state engineering~\cite{krauter11,schindler12}.
Moreover, engineered dissipation can be used to implement digital quantum simulators using Rydberg atoms~\cite{weimer10,weimer11} or trapped ions as demonstrated recently~\cite{barreiro12}.
For a review on open and dissipative systems see Ref.~\cite{mueller12,daley14}.

The study of open quantum many body systems has attracted a lot of interest recently such as dephasing dynamics in interacting quantum systems~\cite{poletti13,bernier13,zi13,schachenmayer14} and also reveals new phenomena such as dissipative phase transitions~\cite{carmichael80,capriotti05,werner05,morrison08,diehl10c,tomadin11,kessler12,sieberer13,dallatorre13}.
Although they share certain features with conventional (quantum) phase transitions~\cite{sachdev99b,kessler12}, extended concepts such as new dynamical universality classes~\cite{sieberer13} offer new fields of research that go beyond the equilibrium understanding of universality. 

We consider an open systems that couples to a Markovian bath in the following.
After tracing out the bath degrees of freedom, the time evolution of the (system) density matrix $\rho$ is given by a master equation in Lindblad form~\cite{gorini76,lindblad76,gardinger04,breuer10} 
\begin{equation}
	\partial_t \rho = i[\rho,\mathcal{H}] + \mathcal{L}[\rho].
	\label{eq:master}
\end{equation}
The first term simply reproduces the von Neumann equation and generates the unitary time evolution.
The interaction with the bath is encoded in the Liouville operator $\mathcal{L}$.
The idea of dissipative state preparation is that $\mathcal{L}$ has a unique and pure dark state $\ket \Omega$ with $\mathcal{L}[\ketbra{\Omega}{\Omega}]=0$.
If this state is also an eigenstate of $\mathcal{H}$ it will be a stationary solution of \Eref{eq:master}.
If, however, the unitary time evolution is not compatible with the dark state of $\mathcal{L}$, the steady-state solution will in general be mixed and determined by the non-trivial interplay of $\mathcal{H}$ and $\mathcal{L}$.
One can raise the question what steady states are realized when varying the microscopic system and bath parameters, i.e. what the dissipative phase diagram is and how the system equilibrates into the steady-state.

In this work, we address this question in the setting of a Bose-Hubbard chain in contact with a superfluid bath. This setup has been introduced and studied in a series of papers~\cite{diehl08,kraus08,diehl10c,tomadin11}.
By suitably chosen dissipators that act on the bonds between two adjacent lattice sites, the unique dark state of the dissipator is a uniform $k=0$ Bose-Einstein condensate (BEC).
Using a matrix product state inspired superoperator renormalization technique~\cite{zwolak04,verstraete04b,prosen10}, we can numerically resolve the real-time evolution of the full interacting quantum system, while
representing the system density matrix as a matrix product operator (MPO).

We then start by studying the interplay of the non-interacting kinetic energy with the bond dissipation. We find -- somewhat surprisingly at first sight -- that in a system with open boundaries, the combined dynamics has a mixed steady state. We then explore the general interplay by including interactions and analyzing the steady states. We also analyze the "unitary" and "dissipative" parts of the particle currents in the steady state,
which mutually compensate each other.
We then compare the correlation functions in the steady state to a Gibbs ensemble with an effective interaction and temperature, raising the question of thermalization.
The dynamical properties of the equilibration process also allows us to access information about the damping spectrum and we address the question of a possible charge-density wave (CDW) instability
raised in a previous work~\cite{tomadin11}.

This paper is organized as follows.
In Section \ref{sec:model}, we briefly review the model of a coupled driven condensate as it is has been investigated previously~\cite{diehl08,kraus08,diehl10c,tomadin11,mueller12} for a translationally invariant setup.
The results of our numerical simulations for the steady state as well as the relaxation dynamics will be presented in Section \ref{sec:resSS}.
Section \ref{sec:concl} contains a concluding summary of our findings.
We include results of the integration of the single particle problem in Appendix~\ref{sec:single} that supplement the results from the previous sections.
A short review of the superoperator algorithm can be found in Appendix~\ref{sec:method}.

\section{Model}
\label{sec:model}
We study a Bose-Hubbard chain coupled to a superfluid bath such that the dissipative process will lock the phase of adjacent sites leading to an exact condensate with off-diagonal long ranged order (ODLRO )
in the dark state.

The unitary dynamics of the bosons in a one dimensional lattice with open boundary conditions of length $L$ (we set the lattice spacing to $a=1$ and work in units where $\hbar = k_\mathrm{B}=1$) is described by the Bose-Hubbard Hamiltonian
\begin{equation}
	\mathcal{H} = -J \sum_{j=1}^{L-1} \left( b_j^\dagger b_{j+1} + \mathrm{h.c.} \right) 
	    +\frac{U}{2} \sum_{j=1}^L n_j (n_j - 1).
	\label{eq:hamiltonian}
\end{equation}
Here, $J$ is the hopping amplitude between nearest-neighbors, $U$ is the on-site interaction, $b_j$ ($b_j^\dagger$) are bosonic annihilation (creation) operators and $n_j$ counts the number of particles per site.
For integer fillings, the system undergoes a Berezinskii-Kosterlitz-Thouless quantum phase transition~\cite{berezinskii71,kosterlitz73} from a strong-coupling Mott insulator to a superfluid with quasi-ODLRO for small $U/J$.
In case of a generic filling, the superfluid is stable for all values of $J/U>0$~\cite{kuehner98,kuehner00,giamarchi04}.

The dissipative part is described by the Liouville operator
\begin{equation}
	\mathcal{L}[\rho] = \kappa \sum_{j=1}^{L-1} \left( c_{j,j+1} \rho c^\dagger_{j,j+1} - \frac{1}{2}\lbrace c_{j,j+1}^\dagger c_{j,j+1}, \rho\rbrace \right),	\label{eq:Liouv}
\end{equation}
with the Lindblad operators given as $c_{j,j+1}=(b_j^\dagger + b_{j+1}^\dagger)(b_{j} - b_{j+1})$ and $\kappa$ denotes a uniform coupling to the bath.
This form of the dissipation has the property that the (unique) dark state of $\mathcal{L}$ is pure and can be cast in the form of a non-interacting BEC condensate wave function, $|\Omega \rangle \propto (\tilde b_{k=0}^\dagger)^N |0\rangle$~\cite{diehl08,kraus08}, where $\tilde b_k^\dagger$ creates a particle at momentum $k$.
This can be readily verified by considering the momentum space representation of the Lindblad operators, $\tilde c_k \sim \sum_q (1+e^{i(q-k)})(1-e^{-iq}) b^\dagger_{q-k} b_q$, whose zero mode is given by $|\Omega \rangle$, i.e. $\tilde c_k |\Omega \rangle =0$. Note that in a uniform chain with periodic boundary conditions, the dark state of $\mathcal{L}$ is also an eigenstate of the kinetic energy.

A physical implementation of this model, as described in Ref.~\cite{diehl08}, is a lattice system immersed into a superfluid bath~\cite{griessner06} in a superlattice between to neighboring sites $j$ and $j+1$.
Reminiscent of dark state laser cooling~\cite{esslinger96,morigi98,griessner06}, a Raman transition couples anti-symmetric states, $b_j-b_{j+1}$, to the bath that acts as a reservoir for Bogoliubov excitations.
The excited state will decay into the symmetric (dark) state, $b_j^\dagger + b_{j+1}^\dagger$, giving rise to a phase locking between adjacent sites.

Despite the presence of a pure dark state, this model features rich physics resulting from the interplay of interaction and dissipation giving rise to a dynamical phase transition and has been studied using linearized equations of motions and mean-field like methods~\cite{diehl08,kraus08,diehl10c,tomadin11}.
In the absence of the $U$-term in the Hamiltonian, the steady state is given by the BEC state $|\Omega \rangle$, as discussed above.
For $d >1$, finite interaction gives rise to an effective temperature that will, for large enough $U$, drive the system into a mixed state for long times where (quasi) ODLRO is lost.
The two regimes are separated by a continuous dynamical phase transition where the condensate fraction exhibits universal scaling~\cite{diehl10c,tomadin11}.
A second feature of the dynamical phase diagram of this system is the appearance of an instability for small momenta in the damping spectrum of $\mathcal{L}$ for large values of $\kappa$~\cite{diehl10c,tomadin11}.
This instability can manifest itself in the appearance of charge density wave ordering that, however, will be only visible on large length scales~\cite{diehl10c,tomadin11}.
In one dimension, phase coherence is lost for any finite temperature since it is expected that interactions will totally prohibit the existence of a dark pure state in the proposed setup~\cite{diehl08}.

\section{Numerical Results}
\label{sec:resSS}
We simulate the real-time dynamics of the system by evolving the system in time according to the full master equation starting from some initial state.
In our simulations, this can either be the ground state of $\mathcal{H}$ at some fixed value of $U/J$ or a thermal Gibbs state $\rho(0)=\exp(-\beta \mathcal{H})$.
In the following section, we will discuss the nature of the steady state reached for sufficiently long times independent of the initial conditions, as well as its short time and transient dynamics.

One important aspect in the system considered here is the choice of open boundary conditions that are reminiscent of a the situation in a realistic implementation in cold atomic systems.
The existence of the unique ground state $|\Omega \rangle$ relies a momentum space representation of the Lindblad operators (see discussion in Section \ref{sec:model} and references therein).
Although this setup in not translationally invariant, the phase locking mechanism between nearest neighbor sites is ought to hold thus we expect $\mathcal{L}$ to have a homogenous dark state with long-range phase coherence, and our numerical results support this expectation.
The kinetic energy term of $\mathcal{H}$, however, can not be minimized on all bond simultaneously in the absence of the bond connecting sites 1 and $L$ as in a setup with, say, periodic boundary conditions.
Thus, the density at the border of the system is depleted and the the ground state is inhomogeneous.
In particular, $|\Omega \rangle$ is not an eigenstate of $\mathcal{H}$ at $U=0$ such that we have the situation of competing unitary and dissipative dynamics even in the absence of interactions.

In order to study the interplay between the two terms in the unitary dynamics and the dissipator separately, we consider two scenarios in \Sref{sec:kindiss}, \ref{sec:curr} and \ref{sec:intdiss}.
First, we study the interplay between the kinetic and dissipative terms ($U=0$) where the boundary effects due to finite system sizes play the most important role.
In the second part the effects of finite interactions $U>0$ with and without kinetic term are studied.
\Sref{sec:thermal} compares our steady state results to a thermal ensemble.

The dynamical properties and the convergence towards the steady state in particular are discussed in \Sref{sec:dyn}.

\begin{figure}[t]
 \begin{center}
  \includegraphics[width=\columnwidth]{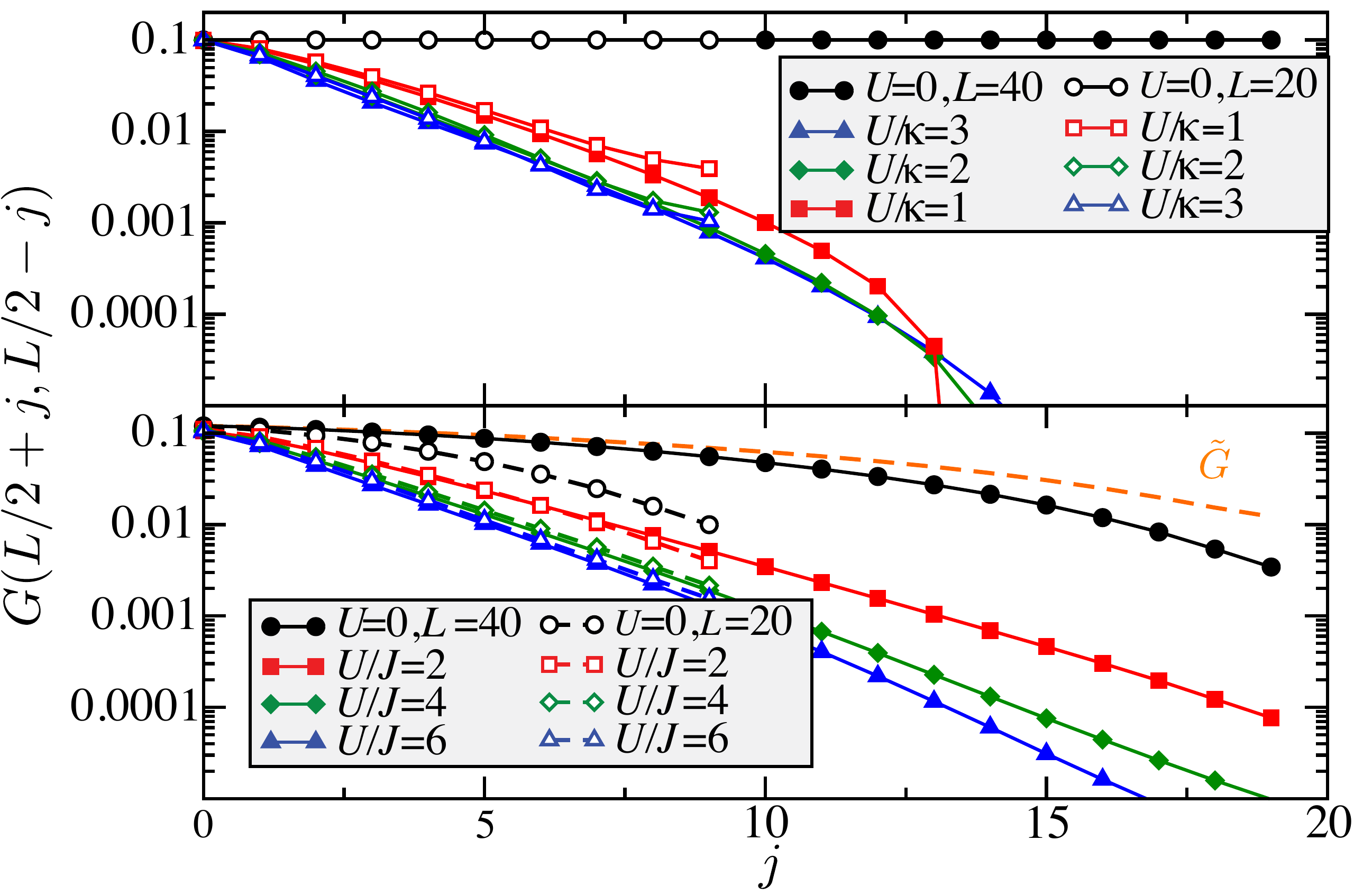}
 \end{center}
\caption{(Color online)
Green's Function $G(L/2-j, L/2+j)$ for the steady-state at filling $n=0.1$ and $L=20$ and $40$.
{\it Upper panel:} Simulation results for $J=0$, i.e. the OBC kinetic energy does not contribute to the time evolution.
{\it Lower panel:} Finite nearest neighbor hopping (here $\kappa/J=2$) suppresses qODLRO even for vanishing interaction.
The orange symbols represent the density normalized Green's function $\tilde G(j,l)$ normalized such that $\tilde G(L/2,L/2) = G(L/2,L/2)$.
The horizontal dashed line denotes $G(j,L/2)=0.1$.
}
\label{fig:GreenSS}
\end{figure}

\subsection{Interplay between kinetic energy  and dissipation in systems with open boundaries}
\label{sec:kindiss}
\begin{figure}[t]
 \begin{center}
  \includegraphics[width=\columnwidth]{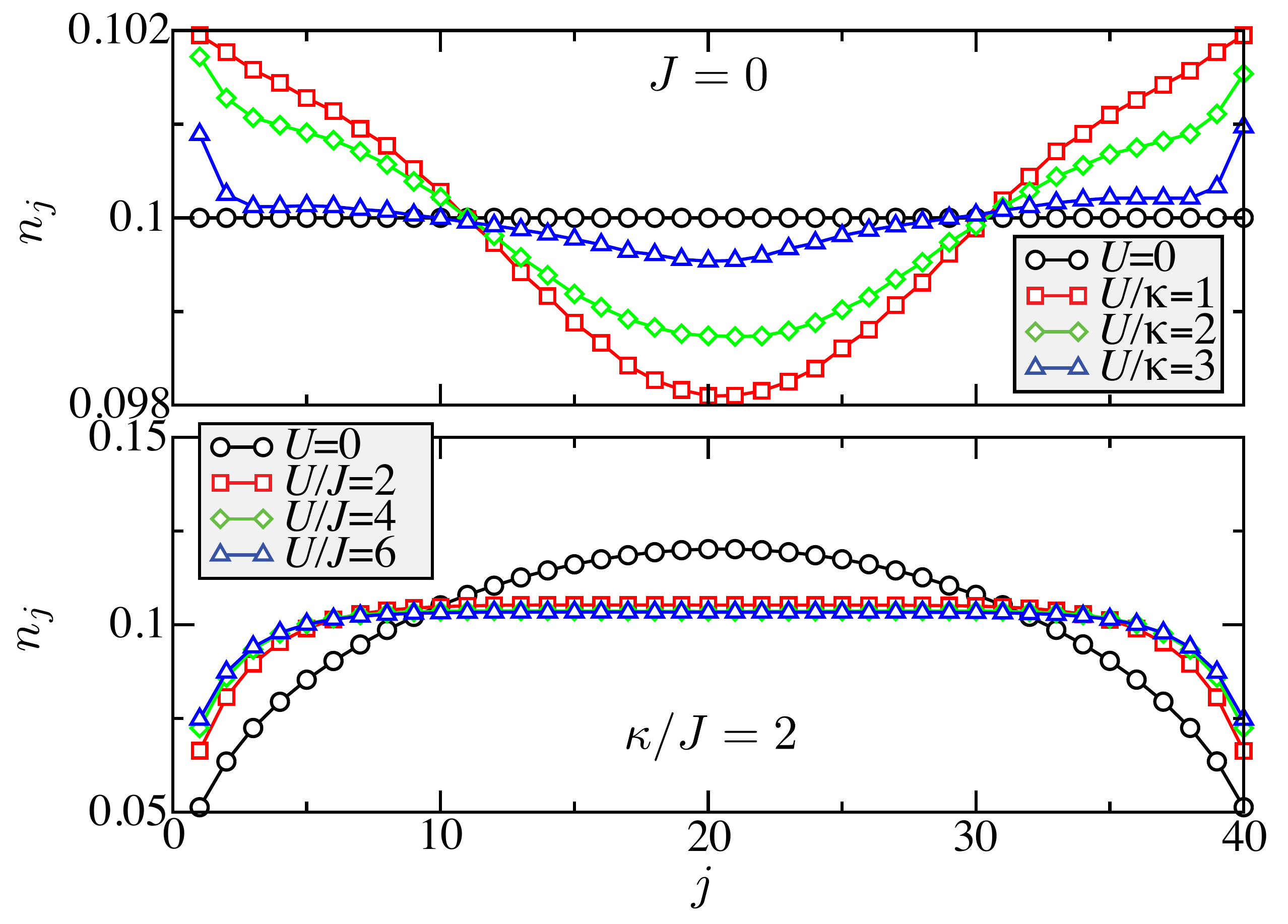}
 \end{center}
\caption{(Color online)
Real space density distribution $n_j$ for the steady state of a chain with $L=40$ sites at filling $n=0.1$ for $J=0$ (upper panel) and $\kappa/J=2$ (lower panel) and different values of the interaction $U$.
}
\label{fig:DensitySS}
\end{figure}

First, we study the steady-state properties by evolving the initial state, chosen as the ground state of $\mathcal{H}$, for long times until the observables, namely the equal time Green's function,
\begin{equation}
G(j,l)=\mathrm{Re} \langle b_j^\dagger b_l \rangle,
\end{equation}
the local particle number $n_j$, the energy and the entropy are converged to their steady state values.
\Fref{fig:GreenSS} and \Fref{fig:DensitySS} show $G(j-L/2,j+L/2)$ and $n_j$ for the steady state at filling $n=0.1$ for various values of the interaction considering the two protocols mentioned above.
The low density allows us to perform our simulations without the introduction of a particle number cut-off and we find that a MPS rank of $\chi=300$ is sufficient for system sizes up to $L=50$.
Here, we use a fourth-order Trotter decomposition with a time step of $\delta t J = 0.03$.
Considering the first protocol where the kinetic term is switched off at $t=0^+$, we find that in the purely dissipative case ($U=0$) the dark state is in fact given by a homogenous state with $G(j,l)=n_j=n$ exhibiting long-range phase correlations.
In the presence of the kinetic energy with open boundary conditions, however, ODLRO is apparently lost in the steady-state, since $G(j,l)$ is suppressed at the boundaries.
The influence of the kinetic term is also reflected in the density profile where the boson occupation is diluted near the boundaries.
Note that the  density redistribution is not the main source for the suppression of $G$.
Also the density normalized Green's function $\tilde G(j,l)=G(j,l)/\sqrt{n_j n_l}$ is suppressed for large distances $|j-l|$, as shown in \Fref{fig:GreenSS}.
In the bulk of the chain one finds a growing interval where $G(j,l) \approx n$ -- boundary effects  thus become less important as $L$ is increased -- such that the bulk for a large enough system looks similar to the pure dark state. 
It is noteworthy that already for the single particle problem that can be integrated directly, as it is discussed in \ref{sec:single}, the system does not approach a pure state as $L \rightarrow \infty$: the purity $F=\tr [\rho^2]$ does not extrapolate to $1$ in the thermodynamic limit but is monotonically decreasing, as shown in \Fref{fig:TrRhoFinal_SingleParticle}, although some observables converge towards those of $|\Omega \rangle$.

\subsection{Unitary and dissipative currents in the steady state}
\label{sec:curr}
\begin{figure}[t]
 \begin{center}
  \includegraphics[width=\columnwidth]{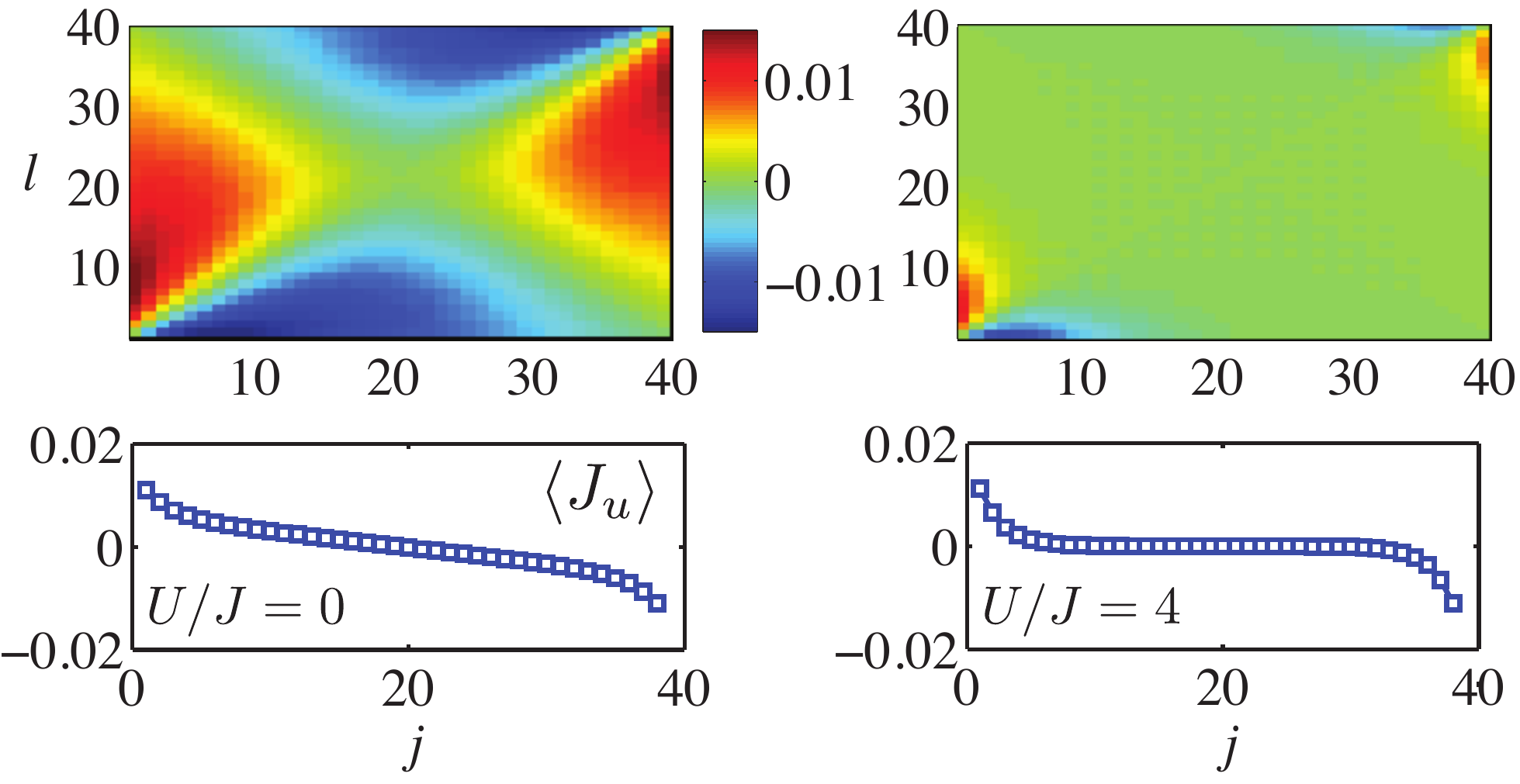}
 \end{center}
\caption{(Color online)
Imaginary part of the Green's function $\langle b_j^\dagger b_l - b_j b_l^\dagger \rangle$ at filling $n=0.1$. 
{\it Upper:} Left (right) panel shows steady state results for $U=0$ ($U/J=4$) and $\kappa/J=2$.
{\it Lower:} Left (right) panel shows the lattice divergence of the unitary current $\langle \mathcal{J}_u \rangle$.
}
\label{fig:Green_Imag}
\end{figure}

To understand the structure of the steady state even further, we turn towards the imaginary part of the Green's function, $I(j,l)=\mathrm{Im} \langle b_j^\dagger b_l \rangle$, whose nearest-neighbor component can be identified with the expectation values of the current operator in the unitary case reading $\langle \mathcal J_{jl} \rangle = -i J \mathcal{Z}^{-1} \tr{ [I(j,l) \rho]}$, where $\mathcal{Z}=\tr[\rho]$ is the partition function.
Here, we have a dissipative term that tries to homogenize the system and drive density to the edges of the system.
This is counteracted by a current emanating from the kinetic term that leads to a flow of particles towards the center of the lattice.
This can be seen by looking at $I(j,l)$ that is shown for the steady state at $L=40$ and different values of the interaction in \Fref{fig:Green_Imag}.
Whereas in the center of the lattice $I$ is almost zero, it acquires a finite values at the boundary giving rise to a current emanating from the kinetic term that is directed towards the center (see lower panel of \Fref{fig:Green_Imag}).
The full expression for the total current can be derived from the equations of motion for the local density $\langle n_j \rangle$ reading
\begin{align}
\begin{split}
\partial_t \langle n_j \rangle &= \tr\left( i[\mathcal{H},n_j] \rho\right) \\
&+ \tr\left( \kappa \sum_{l=l}^L \left[c_{l,l+1}^\dagger n_j c_{l,l+1} - \frac{1}{2} \lbrace n_j, c_{l,l+1}^\dagger c_{l,l+1} \rbrace\right]\rho\right).
\label{eq:EMOn}
\end{split}
\end{align}
\Eref{eq:EMOn} resembles a continuity equation $\partial_t \langle n_j \rangle = \langle \mathrm{div} \mathcal{J} \rangle$ and it is straightforward to show using bosonic commutation relations that the two contributions to the total divergence, originating from the unitary and dissipative term respectively, read
\begin{equation}
	\langle \mathrm{div} \mathcal{J}_u \rangle= -iJ \tr \left(  \left[ b_{j-1}^\dagger b_j - b_{j-1} b_j^\dagger - b_{j+1} b_j^\dagger + b_{j+1}^\dagger b_j \right] \rho \right)
\end{equation}
and
\begin{align}
\begin{split}
\langle \mathrm{div} \mathcal{J}_d \rangle = -\kappa \tr (
 [ 
(&n_j-n_{j+1})(T_{j,j+1}+1) \\+ (&n_{j-1}-n_j)(T_{j-1,j}+1)
 ]
 \rho ).
 \end{split}
\end{align}
Here, $T_{jl}=b_j^\dagger b_l + b_l^\dagger b_j$ is the local kinetic energy.
In particular, one can see clearly that the dissipative part is sensitive towards density gradients that will give rise to a finite dissipative current that has to be canceled by the unitary part in the steady state since $\partial_t \langle n_j \rangle =0$.
For larger values of the interaction (see right panel of \Fref{fig:Green_Imag}), the system realizes the scenario where the current is zero (and the density is flat) except in a small region at the boundaries.
Concentrating on the center of the system, in particular, one can observe a homogenous system with a slightly increased density.

\subsection{Finite interactions}
\label{sec:intdiss}
As interactions are introduced, the Green's function in the steady state decays exponentially with distance and the system thus has a finite correlation length $\xi$, as seen in \Fref{fig:GreenSS}.
This is compatible with the picture drawn by in Ref.~\cite{diehl08} that $U$ will act as an effective temperature for the steady state. This will immediately destroy ODLRO in one dimension.
In particular, the correlation length in the large density limit is predicted to scale as $\xi \propto 1/U$ in the long wavelength limit~\cite{diehl08}.
This scaling is found to hold for $n=1$ (see discussion below).
In the low density limit, $n=0.1$, however, the correlation length seems to decrease very slowly with increasing $U$ due to the diluteness and thus less effective particle interactions.

Thus far, the discussion was mainly focused on the low density limit that can be accessed very effectively because of its small operator space entanglement entropy without the introduction of a local particle number cut-off.
We also consider the case of unit filling ($n=1$) restricting the local Hilbert space to $D=25$ corresponding to a local occupancy $n_j < 5$.
Note that the computational effort of the numerical method scales like $D^3$.
Even for comparably small system sizes up to $L=20$, a bond dimension of a few thousands is required to get accurate results for the time evolution at intermediate times. 

\begin{figure}[t]
 \begin{center}
  \includegraphics[width=\columnwidth]{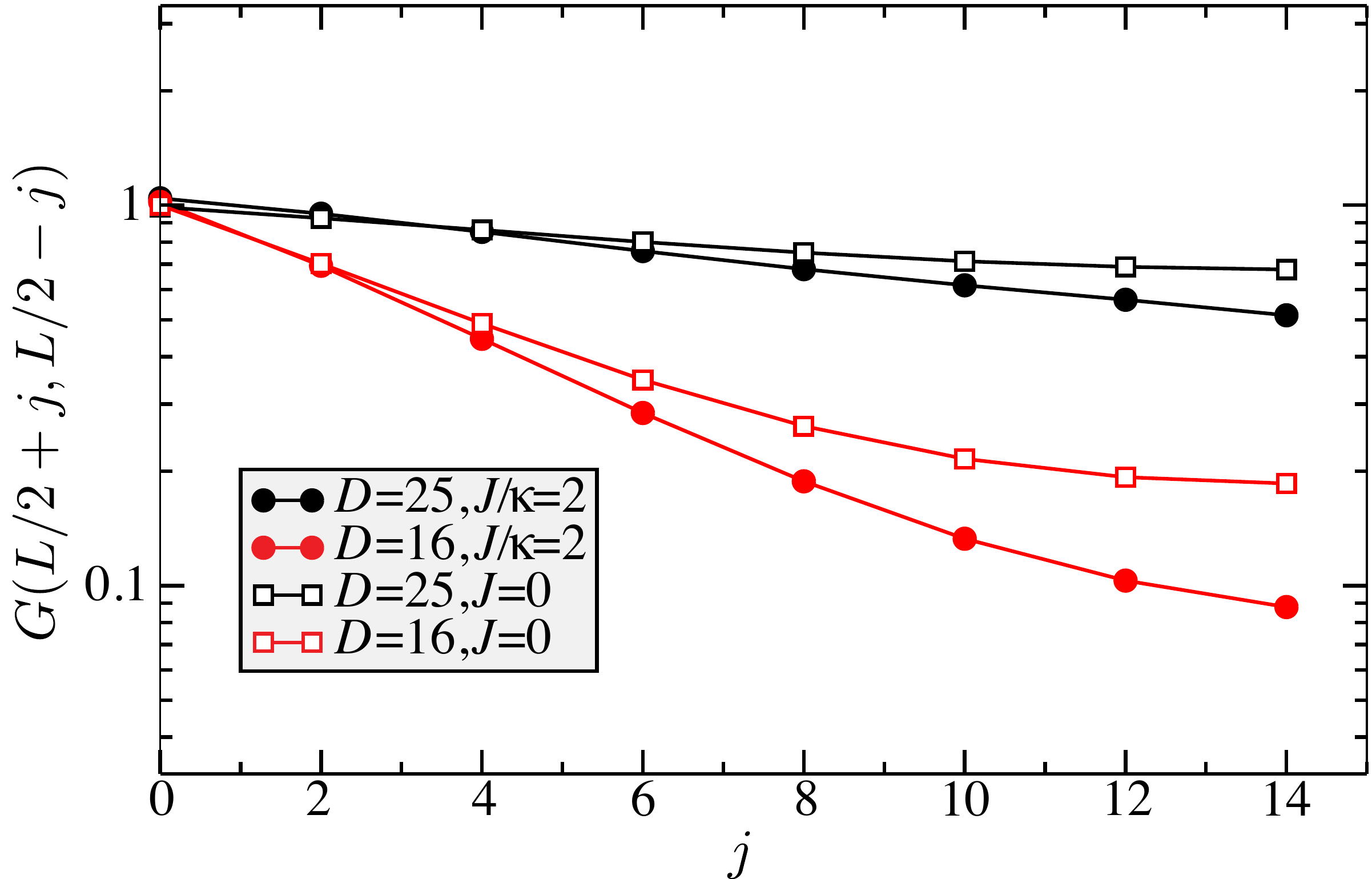}
 \end{center}
\caption{(Color online)
Comparison of the real steady state Green's function $G(j,l)$ for filling $n=1$, different particle number cut-offs $N=\sqrt{D}-1$ and $L=16$ and $U=0$.
The circles denote data obtained with finite $J$ whereas the data for $J=0$ is represented by squares.
The finite particle cut-off acts as an effective interaction an suppresses the Green's function similarly to a finite $U$.
}
\label{fig:CompareGreensFunctionU0}
\end{figure}

\Fref{fig:CompareGreensFunctionU0} shows the Green's function for unit filling for $D=16$ and $25$ for the two aforementioned scenarios.
As $D$ is decreased, the Green's function becomes suppressed even further.
This can be understood by the fact that the particle number cut-off corresponds to an effective hard-core interaction for highly occupied sites translating into an effective temperature that will suppress $G$.
The particle number cut-off is not only a numerical limitation but can also be present in the limit of strong three-body losses, for instance, where triple (and higher) occupation of sites is suppressed~\cite{daley09,schuetzold10}.

\begin{figure}[t]
 \begin{center}
  \includegraphics[width=\columnwidth]{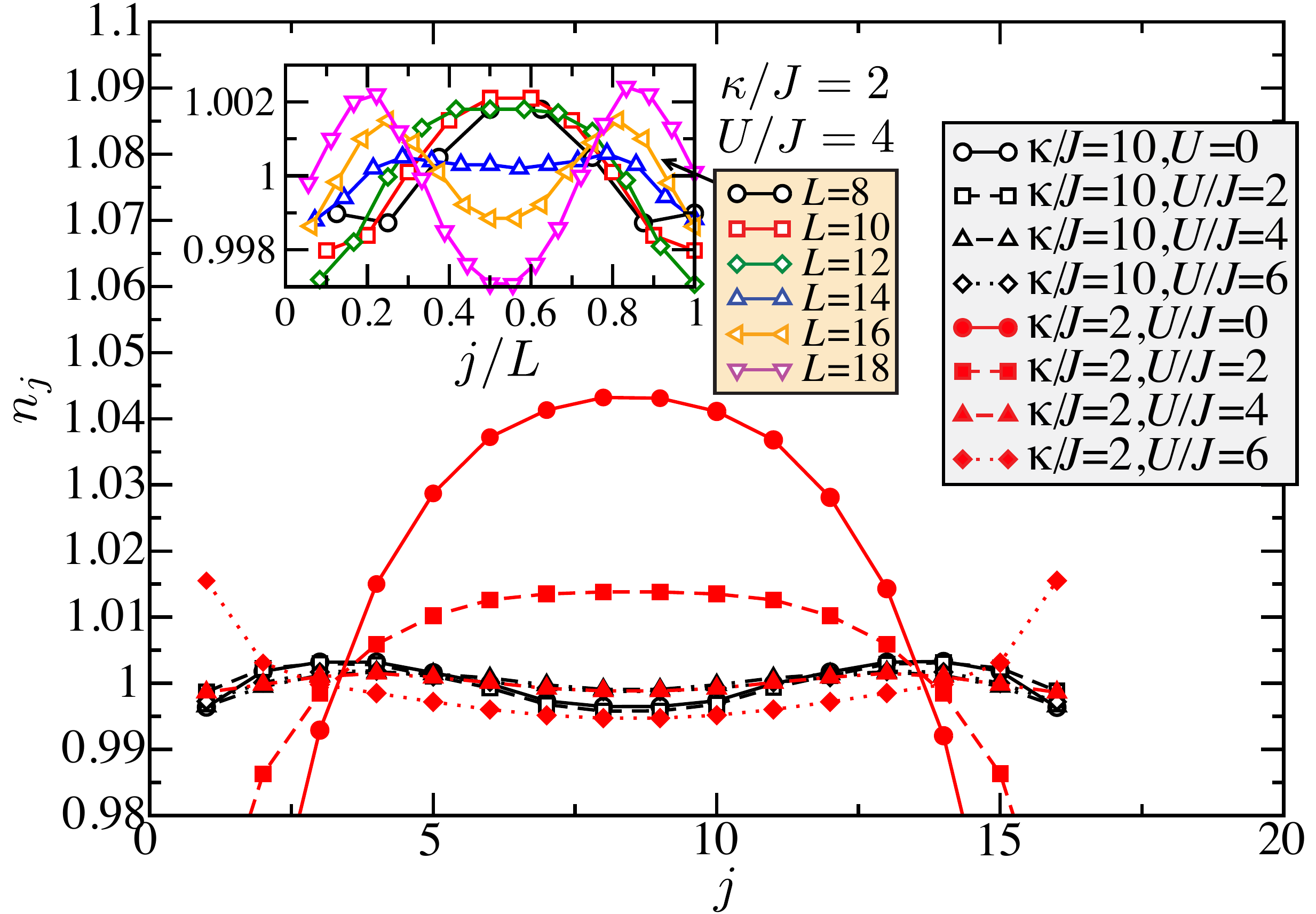}
 \end{center}
\caption{(Color online)
Real space density distribution $n_j$ for the steady state of a chain with $L=16$ sites at filling $n=1$ for $\kappa/J=2$ and $10$.
The inset shows the local density for $U/J=4$ and $\kappa/J=2$ for different system sizes.
}
\label{fig:DensityKappa10Kappa2}
\end{figure}

Although ODLRO is lost for finite interaction for $n=1$ similar to the aforementioned results for $n=0.1$, there is a qualitative difference in the density profile of the steady state as  the density is increased. 
For $U=0$, the density has a dome-like structure. 
As $U$ is increased, however, the density in the center of the system does not remain flat but shows a sinusodial modulation, as can be seen in \Fref{fig:DensityKappa10Kappa2}, that even becomes more pronounced as $\kappa/J$ increases.
The wavelength of this oscillation, however, is proportional to $L$ thus this can not be interpreted as an CDW instability discussed in Section \ref{sec:model} that is expected to have a wavelength on the order of a hundred lattice constants~\cite{diehl10c,tomadin11}.
For small lattices, however, only the lowest momentum mode in the dampening spectrum can become unstable and will lead to a CDW with wavelength $\lambda \propto L$~\cite{tomadin11}, that is a vanishing effect in the thermodynamic limit.
Although that scenario is in principle compatible with our numerical findings, the data is still inconclusive whether a region of possible CDW exists in the steady state phase diagram due to the small system sizes accessible in our simulations.

\begin{figure}[t]
 \begin{center}
  \includegraphics[width=\columnwidth]{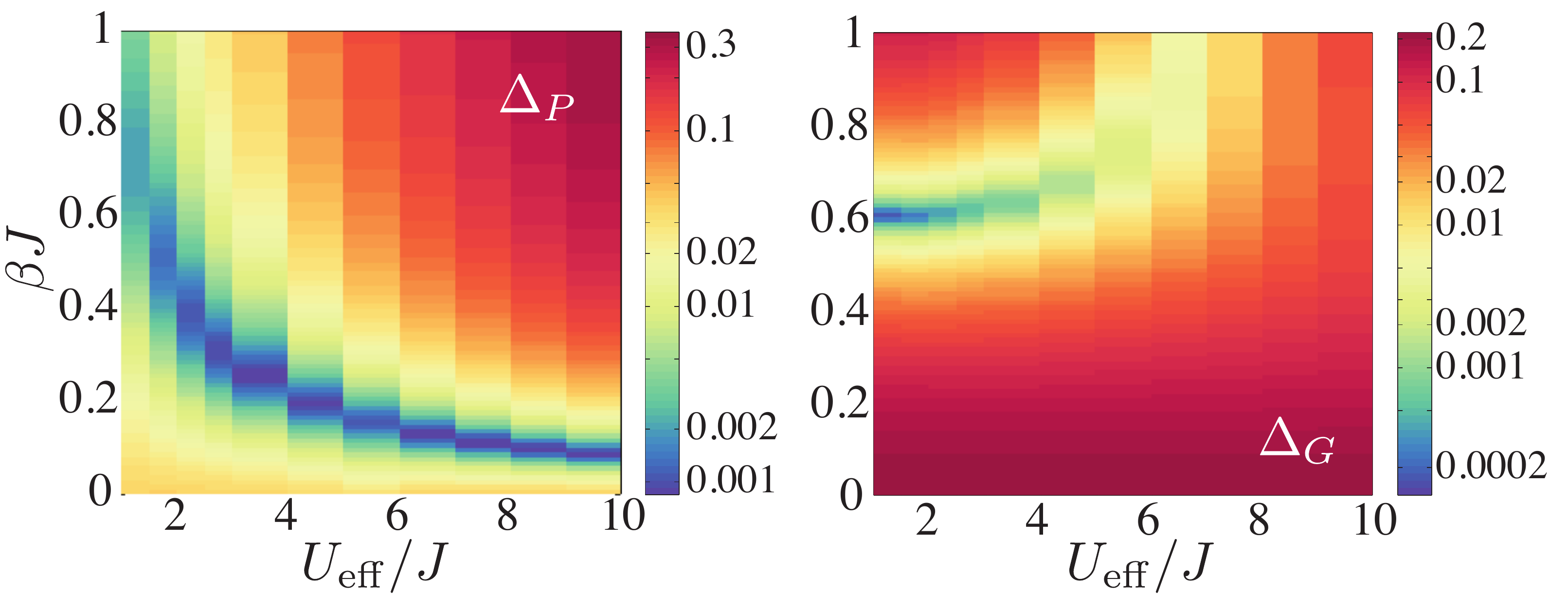}
 \end{center}
\caption{(Color online)
Distances of correlation functions $\Delta_P$(left) and $\Delta_G$ (right) for $n=1$, $L=14$ and $U/J=6$ as a function of effective interaction $U_\mathrm{eff}/J$ and inverse temperature $\beta J$.
}
\label{fig:Delta}
\end{figure}

\subsection{Comparison to Thermal Ensemble}
\label{sec:thermal}
As it has been discussed in Ref.~\cite{diehl08}, the low momentum density matrix looks thermal with an effective temperature $T_\mathrm{eff} \sim U n /2$ that we, however, can not access directly.
Here, we ask the question whether the steady state expectation values of some operators, namely the Green's function and the occupation number projector $P_n=|n\rangle\langle n|$, can be described by a thermal state at some effective interaction $U_\mathrm{eff}$.
To compare the steady-state and thermal expectation values, we use the algebraic distance $\Delta$ of the $G(j,l)$s and $P_n$s respectively, defined as
\begin{equation}
\Delta^2_P=\sum_n \left[ P_n-P_n(U_\mathrm{eff}/J, \beta J) \right]^2
\end{equation}
and
\begin{align}
\begin{split}
\Delta^2_G=\sum_j [ &G'(L/2-j,L/2+j) - \\& G'(L/2-j,L/2+j; U_\mathrm{eff}/J, \beta J)]^2.
\end{split}
\end{align}
Here, $P_n(U_\mathrm{eff}/J, \beta J)=\mathcal{Z}^{-1} \tr[ P_n \exp(-\beta \mathcal{H}(J,U_\mathrm{eff}))]$ denotes the thermal expectation value ($G'(L/2-j,L/2+j; U_\mathrm{eff}/J, \beta J)$ is defined analogously) and we define $G'(j,l)=G(j,l)/G(L/2,L/2)$ in order to reduce the effect of density differences in the bulk of the chain. Note that we are not expecting a thermal state with respect to the generator of the unitary part of the
dynamics, but leave the ratio of interaction to kinetic energy a free parameter. This is plausible, as in the present situation the dissipation broadly acts the same way as the kinetic energy in the bulk.

\begin{figure}[t]
 \begin{center}
  \includegraphics[width=\columnwidth]{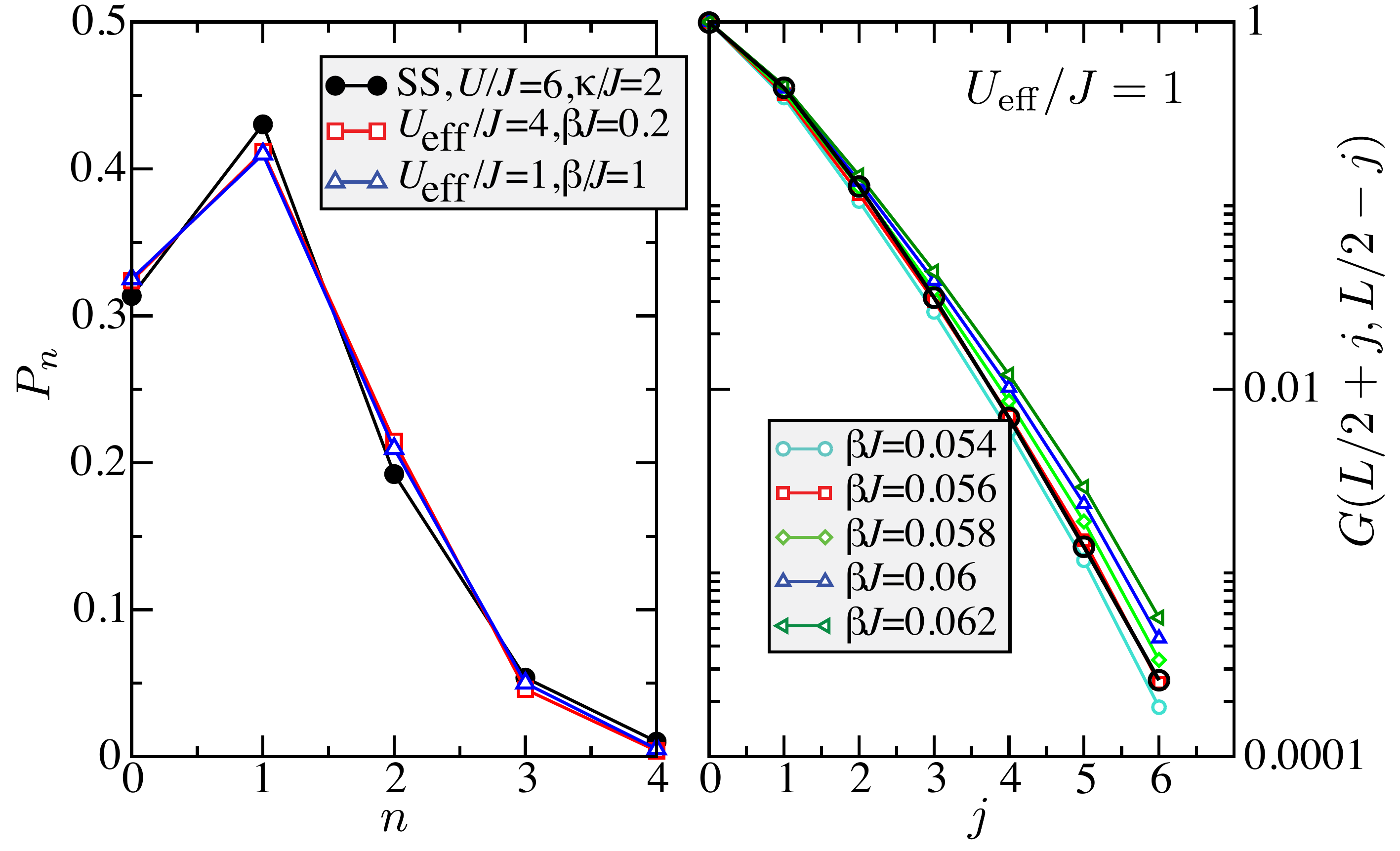}
 \end{center}
\caption{(Color online)
{\it Left:} Particle number projector $P_n$ for the steady state at $L=14$, $n=1$ and $U/J=6$ with thermal results at an effective $U_\mathrm{eff}/J$.
{\it Right:} Comparison of the steady state Green's function for the same parameters compared to the thermal result at $U_\mathrm{eff}/J=3$ at various temperatures.
}
\label{fig:Thermal}
\end{figure}

$\Delta_P$ and $\Delta_G$ show a qualitatively different behavior, as illustrated in \Fref{fig:Delta} for $U/J=6$ at unit filling.
The distances of the occupation number projectors show a broad minimum for $U_\mathrm{eff}/J\sim 1$ to 2 at inverse temperature between $\beta J = 1$ and $0.5$ that extends also to large effective interactions at comparably high temperatures.
The distance of the steady state projectors from the thermal ones can be traced back to an increase of $P_1$ in the steady state, as shown in the left panel of \Fref{fig:Thermal} that dominates $\Delta_P$.

The agreement of the Green's functions, illustrated by $\Delta_G$ in the right panel if \Fref{fig:Delta}, is good only in a small temperature window around $\beta J \sim 0.6$ and $U_\mathrm{eff}/J \lesssim 2$.
A direct comparison for some effective interaction and temperatures is presented in the right panel of \Fref{fig:Thermal}.
This reveals that the nearest- and next-to-nearest neighbor Green's function can be matched quite well to a thermal ensemble.
For large distances -- they only have a small influence on $\Delta_G$ due to the strong decay of $G'$--, deviations become significant and might also be a result of the strong suppression of $G'$ at the boundaries.

\subsection{Dynamical Properties and Convergence towards the Steady State}
\label{sec:dyn}
\begin{figure}[t]
 \begin{center}
  \includegraphics[width=\columnwidth]{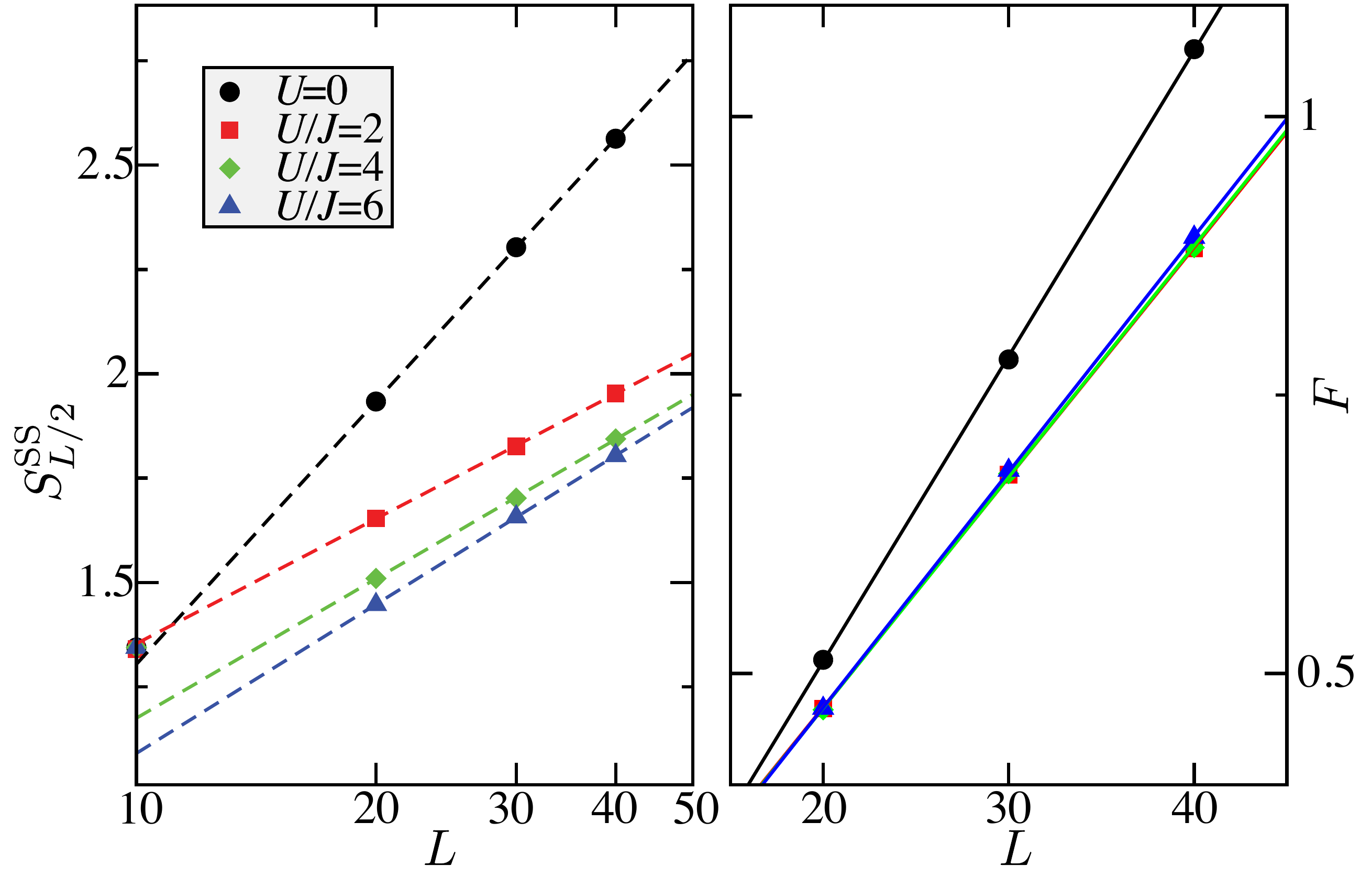}
 \end{center}
\caption{(Color online)
Steady-state operator space entropy (left) and bipartite fluctuations $\mathcal{F}$ for a block size of $L/2$. 
The dashed lines in the left panel denote fits to $a\log L+b$ where the full lines on the right hand side are linear fits to $AL+B$.
}
\label{fig:maxEntrK2}
\end{figure}

Finally, we analyze the operator space entanglement entropy, defined in \Eref{eq:defS}, for a bipartition at $L/2$ in the steady state.
A finite-size extrapolation of $S^\mathrm{ss}_{L/2}$, shown in \Fref{fig:maxEntrK2}, reveals a logarithmic scaling.
Unlike the case of a critical system in one dimension where the logarithmic scaling originates from corrections to the area law~\cite{eisert10b}, particle number conservation can impose a constraint on the density matrix that translates into a finite MPS rank.
Consider for instance the density matrix representing an infinitely hot state.
For a sector of fixed particle number $N$ it can formally be described by applying the projector $\mathcal{P}_N$ to the $N$-particle subspace, $\rho_N = \mathcal{P}_N \rho_\infty \mathcal{P}_N = \mathcal{P}_N$, whose operator space entanglement scales like $\log L$~\cite{muth11}.
In addition, we calculate the bipartite fluctuations $\mathcal{F}(l)=\langle N_l^2\rangle - \langle N_l \rangle^2$,~\cite{song12}, where $N_l = \sum_{j<l} n_j$.
They exhibit linear scaling as it is predicted for a thermal state~\cite{song12,rachel12}.

\begin{figure}[t]
 \begin{center}
  \includegraphics[width=\columnwidth]{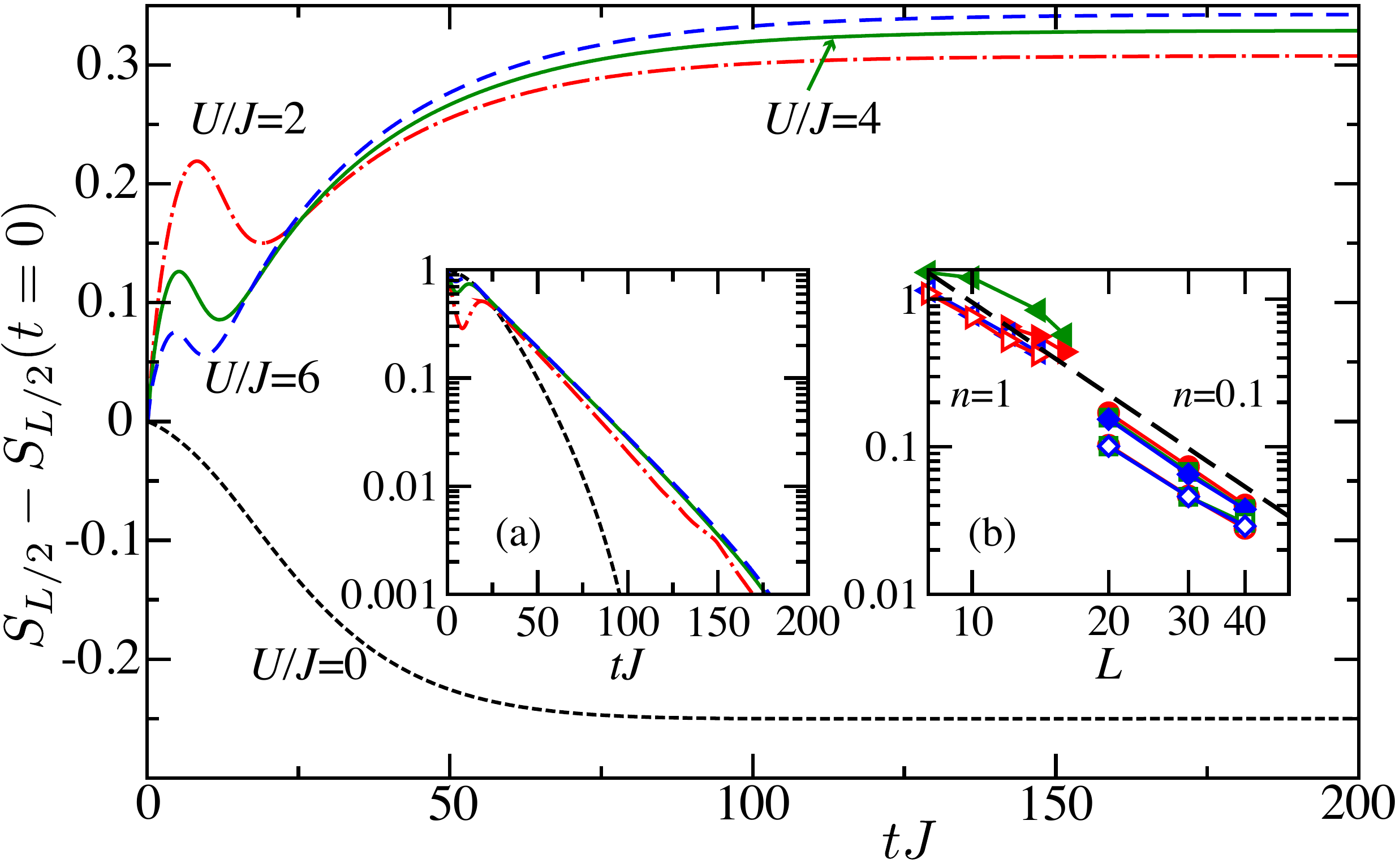}
 \end{center}
\caption{(Color online)
Increase of operator space entanglement entropy $S_{L/2}$ with respect to the initial state for $L=40$, $n=0.1$ $\kappa/J=2$ and different values of the interaction strength.
{\it Inset} (a): The figure shows $1-[S_{L/2}(t)-S_{L/2}(t=0)]/[S_{L/2}^{ss}-S_{L/2}(t=0)]$ in order to illustrate the exponential convergence of the entropy towards the steady state values for the data shown in the main panel.
{\it Inset} (b): Relaxation rate $\alpha$ of the long-time behavior of $S$.
The $n=0.1$ data corresponds to $\kappa/J=2$ (full symbols) with interaction parameters $U/J=2$ (red circles), 4 (green squares) and 6 (blue diamonds) and $J=0$ (hollow symbols) with $U/\kappa=1$ (red circles), 2 (green squares) and 3 (blue diamonds).
For $n=1$, we show data for $J=0$ and $U/\kappa=2.5$ (blue triangles) and 3 (red triangles).
They obey a power law scaling $\alpha \sim L^{-2}$, indicated by a dashed line, compatible with the low-momentum damping spectrum of $\mathcal{L}$.
}
\label{fig:Entropies}
\end{figure}

The convergence towards the steady state is governed by the damping spectrum, the real part of the eigenvalues, of the Liouville operator $\mathcal{L}$.
We find that the observables, for instance the entropy shown in \Fref{fig:Entropies}, convergence exponentially to their steady state values.
The decay constant $\alpha$ is obtained by fitting the entropy to the form $S_{L/2}(t) = C e^{-\alpha t \kappa} + S^\mathrm{ss}$.
$\alpha$ obeys a power law and scales like $\kappa L^{-2}$ -- \Fref{fig:Entropies}(b) shows the decay rate for different values of the couplings and fillings for different fillings and interactions (see also \Fref{fig:ConvTrRho_SingleParticle} for the single particle case).
This is consistent with the results obtained by analyzing the linearized equations of motions~\cite{diehl10c,tomadin11} showing that, albeit the possibility of a CDW instability, the damping spectrum for small momenta $q$ has the form $\kappa q^2$.
In particular, this system does not have a dissipative gap and the convergence time diverges in the thermodynamic limit.

\begin{figure}[t]
 \begin{center}
  \includegraphics[width=\columnwidth]{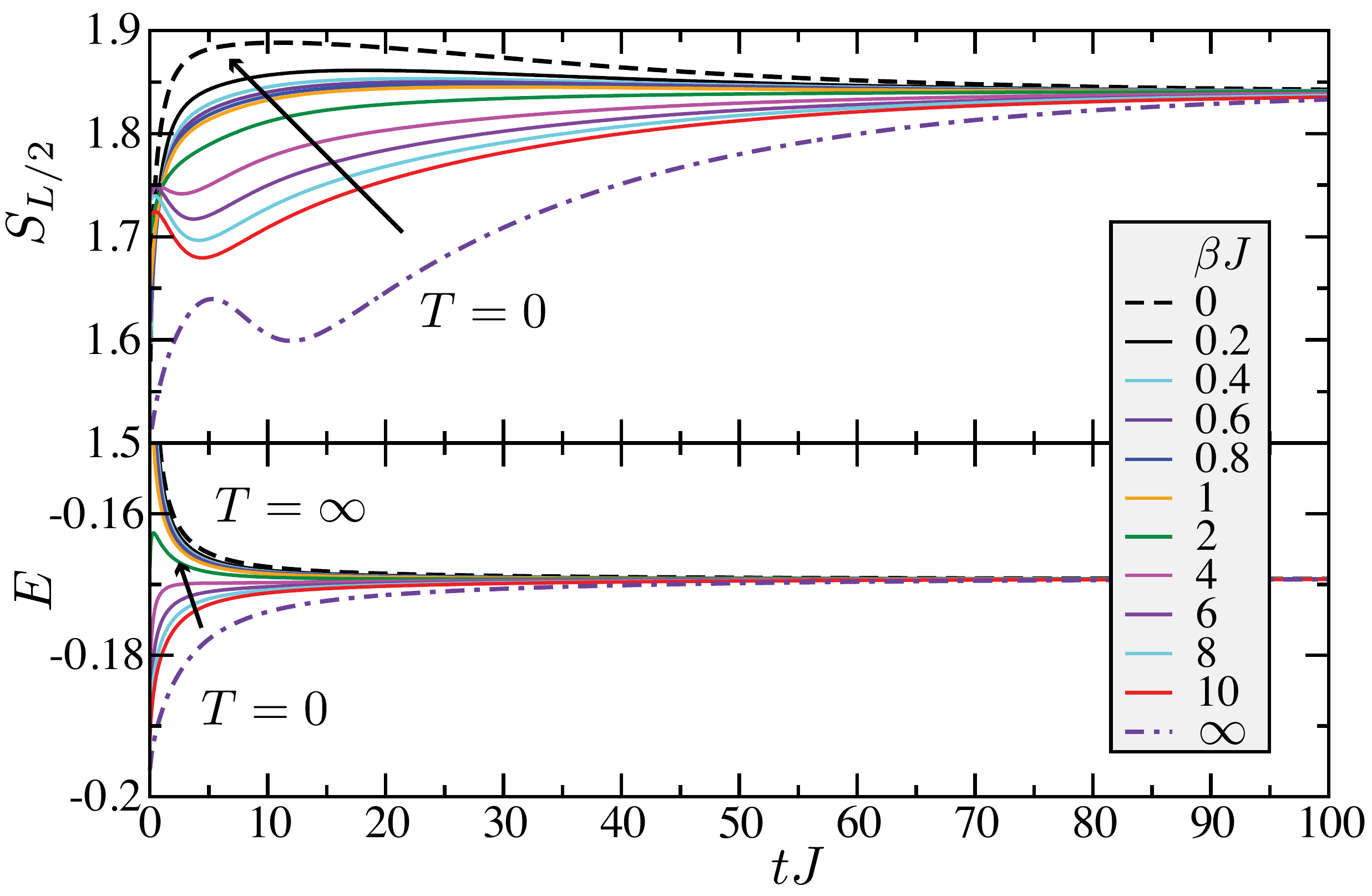}
 \end{center}
\caption{(Color online)
Time evolution of the operator space entanglement (top) and energy (bottom) for $U/J=4$ and $\kappa/J=2$ for different initial thermal density matrices.
The upper dashed line correspond to the $T=0$ and the lower dash-dotted line to the $T=\infty$ initial state whereas the full lines correspond to temperatures $\beta J=0.2$, 0.4, 0.6, 0.8, 1, 2, 4, 6, 8 and 10.
The arrows indicate increasing temperatures.
}
\label{fig:EntropyThermal}
\end{figure}

Although the initial density matrix does not affect the steady state of the system, the transient dynamics differs.
Thus far we used the ground state of $\mathcal{H}$ as initial density matrix but the superoperator framework allows us to start from arbitrary mixed states.
In particular, we consider thermal (Gibbs) initial states at different temperatures.
Whereas the time evolution starting from the ground state shows only a mild increase in entropy for small times, the short time dynamics of a thermal initial states a high temperatures, exemplified in \Fref{fig:EntropyThermal} (see also \Fref{fig:Entropies}), shows a strong initial increase of $S_{L/2}$ even exceeding its steady state value.
The convergence of the energy and the entropy to their large time value, on the other hand, is significantly faster for the large-$T$ states.
As the steady state for finite interactions only has very short ranged correlations, to some extent resembling those of a high-$T$ state as discussed in \Sref{sec:thermal}, quantum correlation in the ground state have to be diminished by the dissipator.
Hence, it can be understood that the large-$T$ states will eventually converge faster as they are more "classical" than low-$T$ or ground states.

\section{Conclusion}
\label{sec:concl}
We applied the superoperator renormalization algorithm to a Bose-Hubbard chain with engineered bond dissipation with a BEC dark state of the Liouvillian.
In contrast to a translationally invariant setup, frustration of the kinetic energy leads to a non-trivial interplay with the dissipator that drives the system into a mixed state.
As a consequence, the Green's function is suppressed at the boundaries but the bulk for large enough system is similar to the pure dark state.
Although a possible CDW instability has eluded itself from our calculations -- possibly due to the small system sizes accessible with our method -- a possible precursor of this phenomenon in terms of a long-wavelength density modulation was found.
For the interacting system, ODLRO is lost completely and the correlation functions compare quite well to thermal states at some effective interaction.

Apart from steady-state phenomena, we present data for the time evolution from which we probe the low-wavelength nature of the damping spectrum and confirm a closing of the dissipative gap for large system sizes.
Although the damping in the long-time limit is solely determined by the spectrum of the superoperator, the built-up and convergence of operator space entropy for thermal initial states shows intriguing properties such that the Gibbs initial systems at $T=0$ and $T=\infty$ systems provide upper and lower bounds for the operator space entanglement for thermal initial states.

The presented study has direct consequences for possible experimental realizations as they illustrate how boundary effects affect the nature of the steady-state and can lead to unwanted heating.
The convergence to the steady-state, on the other hand, improves for mixed initial states at intermediate temperatures highlighting the feature that an initial state preparation is not needed but (almost) each initial system will be driven to the same steady-state.

\section*{Acknowledgments}
We acknowledge discussions with S. Diehl and B. Kraus.
This work was supported by the Austrian Science Fund (FWF) through the SFB FoQuS (FWF Project No. F4018-N23) and the Austrian Ministry of Science BMWF as part of the UniInfrastrukturprogramm of the Research Platform Scientific Computing at the University of Innsbruck.

\appendix

\section{Integration of the Single Particle Problem}
\label{sec:single}
We integrate the equations of motion for a single boson on a chain of length $L$ using a fourth-order Runge Kutta integrator.
This simple integration gives us some insight into the convergence and nature of the steady state on the single particle level and supplement the findings for the many-body problem discussed in this paper.

\begin{figure}[t]
 \begin{center}
  \includegraphics[width=\columnwidth]{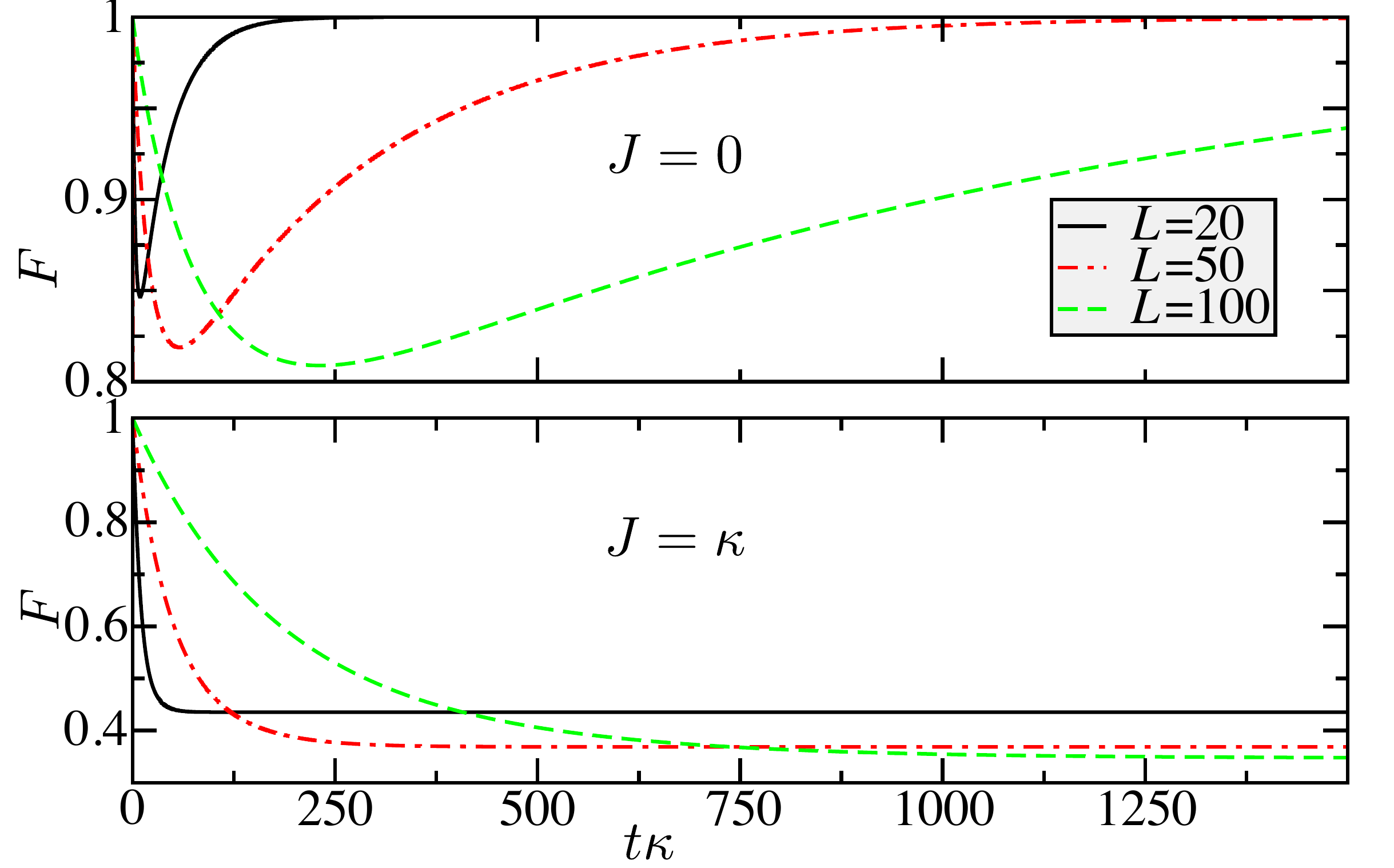}
 \end{center}
\caption{(Color online)
$F=\tr \rho^2$ for the single particle problem for $J=0$ (upper panel) $J=\kappa$ for different systems sizes.
}
\label{fig:TrRhoSq_J0}
\end{figure}

\begin{figure}[t]
 \begin{center}
  \includegraphics[width=\columnwidth]{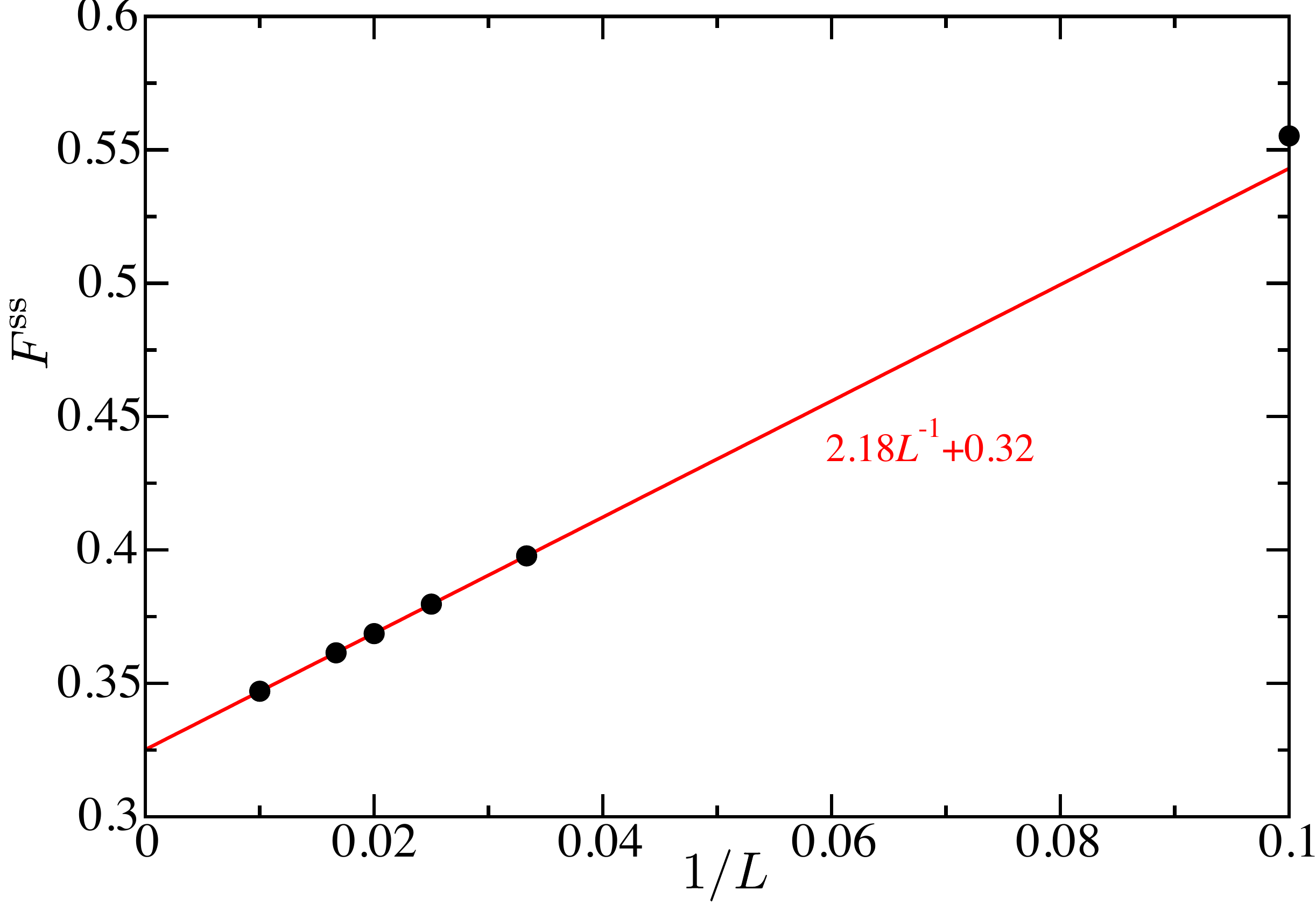}
 \end{center}
\caption{(Color online)
Finite-size extrapolation of $F=\tr \rho^2$ for the single particle problem at $J=\kappa$.
}
\label{fig:TrRhoFinal_SingleParticle}
\end{figure}

First, \Fref{fig:TrRhoSq_J0} shows the real time evolution of purity $F=\tr \rho^2$ starting from the ground-state of $\mathcal{H}$.
If the density matrix evolves purely dissipative ($J=0$), the system will be mixed in the transient regime and eventually converge towards the pure state $|\Omega\rangle$.
The competition of kinetic energy and the dissipator, however, drives the system into a mixed state for long times.
The steady state value of $F$ can be extrapolated to a finite value in the thermodynamic limit (see \Fref{fig:TrRhoFinal_SingleParticle}), i.e. the state is mixed even for $L\rightarrow \infty$ and $F^\mathrm{ss} \rightarrow 0.32$.

\begin{figure}[t]
 \begin{center}
  \includegraphics[width=\columnwidth]{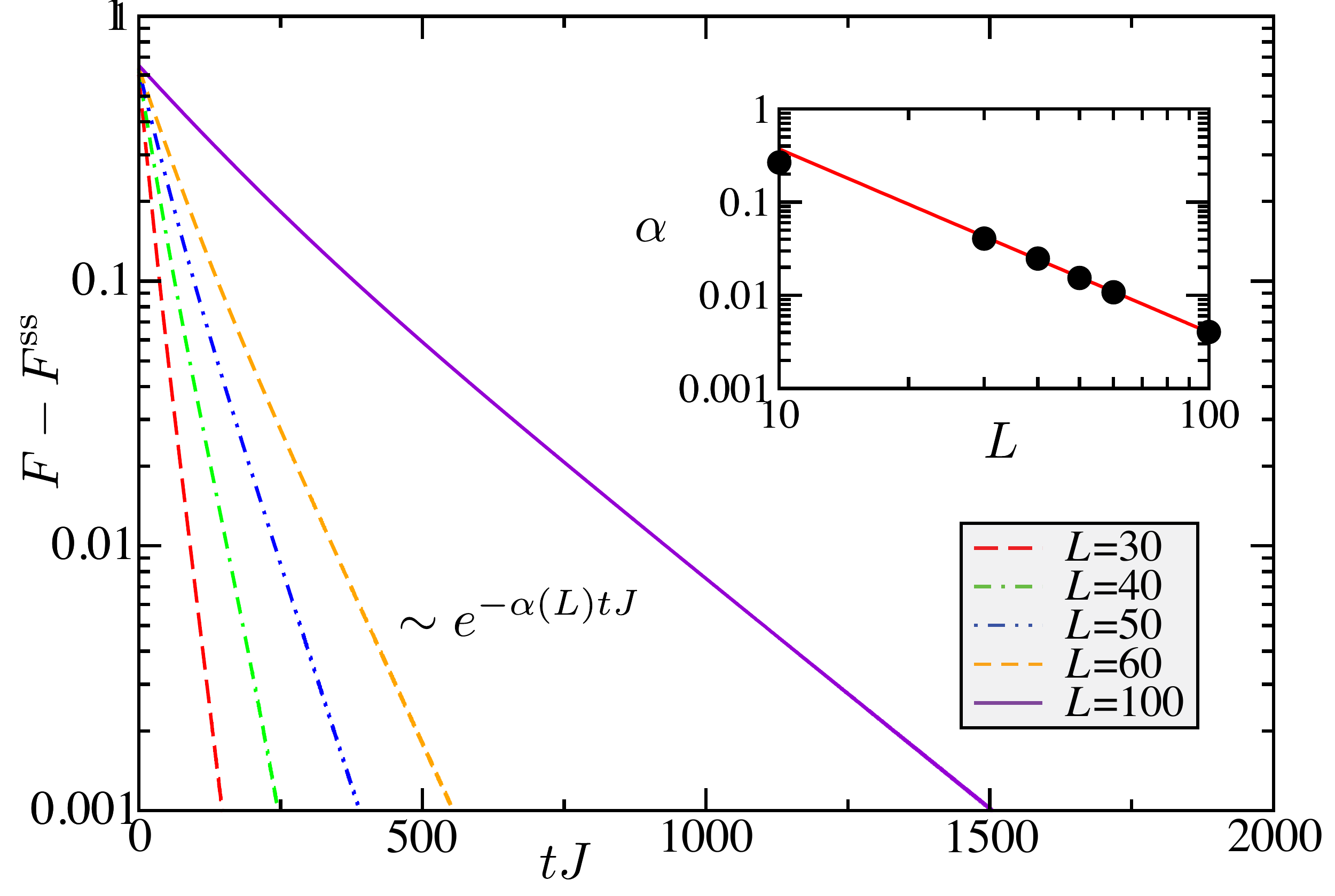}
 \end{center}
\caption{(Color online)
Convergence of $F=\tr \rho^2$ for the single particle problem at $J=\kappa$.
{\it Inset:} Power law behavior of the exponent $\alpha=\tau^{-1}\propto L^{-2}$.
}
\label{fig:ConvTrRho_SingleParticle}
\end{figure}

The convergence towards the steady state is exponential, $F(t)-F^\mathrm{ss} \propto \exp(-\alpha t \kappa)$, as \Fref{fig:ConvTrRho_SingleParticle} clearly shows, where the damping rate $\alpha$ scales with system size as $\alpha \propto L^{-2}$.
This can be attributed to the dampening spectrum that is found to be of the form $\kappa q^2$~\cite{diehl08,diehl10c} for small momenta $q$, where the smallest momentum accessible is proportional to $1/L$ thus leading to a smallest damping rate proportional to $\kappa/L^2$.

The results obtained for the single particle problem are compatible with the MPS simulations including many-particle effects and interactions.

\section{Superoperator Renormalization Group}
\label{sec:method}
We simulate the time evolution of the density matrix governed by the quantum master equation in \Eref{eq:master} using a generalization~\cite{zwolak04} of Time Evolving Block Decimation (TEBD) algorithm~\cite{vidal03,vidal04}.
Within the superoperator renormalization scheme, the density matrix for a chain with $L$ lattice sites and open boundary conditions,
\begin{equation}
\rho=\sum_{i_a, j_a=1}^d c_{i_1, i_2, ... i_L,j_1, j_2, ..., j_L} || i_1, i_2, ..., i_L;j_1, j_2, ,...,j_L \rangle \rangle
\end{equation}
 is represented by a MPS in an enlarged Hilbert space of dimension $D=d^2$, where $d$ is the size of the local Hilbert space $\mathbb{H}$, by consecutive singular value decompositions of the tensor $c_{i_1, i_2, i_3,... ,j_1, j_2, j_3,...}$.
 Thus, we recover the Vidal representation of $\rho$,
 \begin{align}
 \begin{split}
  \rho=\sum_{i_a, j_a=1}^d \sum_{\alpha,\beta,...,\gamma=1}^\chi
  B^{[1] i_1,j_1}_{1,\alpha} \lambda^{[1]}_\alpha B^{[2] i_2,j_2}_{\alpha,\beta} \lambda^{[2]}_\beta ... \\
  \lambda^{[L-1]}_{\gamma} B^{[L] i_L,j_L}_{\gamma,1} 
  || i_1, i_2, ..., i_L;j_1, j_2, ,...,j_L \rangle \rangle.
  \label{eq:rhoMPO}
  \end{split}
 \end{align}
 Here, $|| i_1, i_2, ..., i_L;j_1, j_2, ,...,j_L \rangle \rangle = \bigotimes_{a=1}^L |i_a\rangle\langle j_a| $ are the basis states for the density matrix in the product Hilbert space $\mathbb{H}^{\otimes L} \otimes \mathbb{H}^{\otimes L}$.
The Schmidt spectrum $\lbrace \lambda^{[l]}_\alpha \rbrace$ is truncated keeping only the largest $\chi$ Schmidt values.
This provides a faithful approximation in terms of MPS if the Schmidt spectrum decays fast enough~\cite{zwolak04}.
The equation of motion for the density matrix is integrated using a Suzuki-Trotter decomposition of the time evolution superoperator.
Expectation values with respect to $\rho$ of some operators are calculated using the standard form $\langle O \rangle = \mathcal{Z}^{-1} \tr[ \rho \hat O]$, where $\mathcal{Z}=\tr \rho$ is the partition function.
For convenience, we will drop the $\langle \cdot \rangle$ in this article.

We use imaginary time propagation to prepare our system in the ground state of $\mathcal{H}$.
In the same manner, thermal states can be obtained starting the imaginary time propagation with the infinitely hot state.

In analogy to the entanglement entropy in pure states, the operator space entanglement entropy~\cite{prosen07} of a bipartition $A$ of size $l$, $S_l$, can be obtained from the Schmidt spectrum as
\begin{equation}
S_l=-2\sum_\alpha (\lambda^l_\alpha)^2 \log \lambda^l_\alpha.
\label{eq:defS}
\end{equation}
For a pure state $\rho=|\Psi\rangle \langle \Psi|$, $S_l$ is twice the von-Neumann entropy $S^\mathrm{vN}_l=-\sum_\alpha \tau^{[l]}_\alpha \log \tau^{[l]}_\alpha$ of $|\Psi \rangle$, where $\lbrace \tau^{[l]}_\alpha \rbrace$ are the eigenvalues of the reduced density matrix $\tr_B \rho$ and $B$ denotes the complement of block $A$~\cite{znidaric08}.

\bibliographystyle{apsrev4-1}

%

\end{document}